\documentclass[12pt,preprint]{emulateapj}

\usepackage[bf,small]{caption}

\newcommand{\microns}{$\mu$m}

\def\ga{\mathrel{\hbox{\rlap{\hbox{\lower4pt\hbox{$\sim$}}}\hbox{$>$}}}}
\def\la{\mathrel{\hbox{\rlap{\hbox{\lower4pt\hbox{$\sim$}}}\hbox{$<$}}}}

\def\msun{$M$\mbox{$_{\normalsize\odot}$}}

\def\kms{\,km~s$^{-1}$}

\def\arcsec{$^{\prime \prime}$}

\def\hii{H\,{\sc ii}}
\def\rgc{$R_{\rm GC}$}
\def\phigc{$\phi_{\rm GC}$}
\def\halpha{H$\alpha$}

\slugcomment{}

\shorttitle{The metallicity of the Scutum RSG Clusters}
\shortauthors{Davies et al.}

\begin{document}


\title{Chemical abundance patterns in the inner Galaxy: \\ the Scutum Red
Supergiant Clusters}


\author{Ben Davies\altaffilmark{1}, Livia Origlia\altaffilmark{2},
Rolf-Peter Kudritzki\altaffilmark{3}, Don F.\ Figer\altaffilmark{1},
R.\ Michael Rich\altaffilmark{4}, Francisco Najarro\altaffilmark{5},
Ignacio Negueruela\altaffilmark{6}, J.\ Simon Clark\altaffilmark{7} }

\affil{$^{1}$Chester F.\ Carlson Center for Imaging Science, Rochester
Institute of Technology, 54 Lomb Memorial Drive, Rochester NY, 14623,
USA} 
\affil{$^{2}$INAF - Osservatorio Astronomico di Bologna, via
Ranzani 1, 40127 Bologna, Italy} 
\affil{$^{3}$Institute for Astronomy, University of Hawaii, 2680
Woodlawn Drive, Honolulu, HI, 96822, USA} 
\affil{$^{4}$Department of Physics and Astronomy, UCLA, 430 Portola
  Plaza, Box 951547, Los Angeles, CA 90095-1547, USA}
\affil{$^{5}$Instituto de Estructura de la Materia, Consejo Superior
  de Investigaciones Cientificas, Calle Serrano 121, 28006 Madrid,
  Spain.}
\affil{$^{6}$Departamento. de F\'{i}sica, Ingenier\'{i}a de Sistemas y
  Teor\'{i}a de la Se\~{n}al, Universidad de Alicante, Apdo. 99, E03080
  Alicante, Spain} 
\affil{$^{7}$Department of Physics \& Astronomy, The Open University,
  Walton Hall, Milton Keynes, MK7 6AA, UK}

\begin{abstract}
The location of the Scutum Red-Supergiant (RSG) clusters at the end of
the Galactic Bar makes them an excellent probe of the Galaxy's secular
evolution; while the clusters themselves are ideal testbeds in which
to study the predictions of stellar evolutionary theory. To this end,
we present a study of the RSGs' surface abundances using a combination
of high-resolution Keck/NIRSPEC H-band spectroscopy and spectral
synthesis analysis. We provide abundance measurements for elements C,
O, Si, Mg, Ti, and Fe. We find that the surface abundances of the
stars studied are consistent with CNO burning and deep, rotationally
enhanced mixing. The average $\alpha$/Fe ratios of the clusters are
solar, consistent with a thin-disk population. However, we find
significantly sub-solar Fe/H ratios for each cluster, a result which
strongly contradicts a simple extrapolation of the Galactic
metallicity gradient to lower Galacto-centric distances. We suggest
that a simple one-dimensional parameterization of the Galaxy's
abundance patterns is insufficient at low Galactocentric distances, as
large azimuthal variations may be present. Indeed, we show that the
abundances of O, Si and Mg are consistent with independent
measurements of objects in similar locations in the Galaxy. In
combining our results with other data in the literature, we present
evidence for large-scale ($\sim$kpc) azimuthal variations in
abundances at Galacto-centric distances of 3-5\,kpc. While we cannot
rule-out that this observed behaviour is due to systematic offsets
between different measurement techniques, we do find evidence for
similar behaviour in a study of the barred-spiral galaxy NGC~4736
which uses homogeneous methodology. We suggest that these azimuthal
abundance variations could result from the intense but patchy star
formation driven by the potential of the central bar.
\end{abstract}


\section{Introduction} \label{sec:intro}
Massive young clusters are powerful natural laboratories with which to
study many different aspects of astrophysics. They contain large
numbers of coeval stars with similar initial chemical
compositions. Their youth and large initial masses mean that they
contain many massive stars, and hence are intrinsically very bright
objects. Thus, such clusters can be used to make critical tests of
stellar evolution, while they also trace recent star-forming history
and local abundance patterns.

Two such clusters exist in the constellation of Scutum, at distances
of $\sim$6\,kpc. They are among the most massive young clusters in the
Galaxy, with initial masses in excess of 10$^{4}$\msun. Their ages are
tuned in such a way that their most massive remaining stars are in the
Red Supergiant (RSG) phase, and together these clusters contain
$\sim$20\% of all known RSGs in the Galaxy -- cluster 1 (hereafter
RSGC1) contains 14 RSGs, while cluster 2 (RSGC2) contains 26
\citep[][hereafter F06, D07, D08]{Figer06,RSGC2paper,RSGC1paper}.

The clusters are located in the Galactic Plane, and have high visual
extinctions ($A_{V} \approx 25$ and 16 for RSGC1 and RSGC2
respectively). However, the fluxes of the RSGs peak at
$\sim$1\microns, making them extremely bright in the near-IR where the
extinction is reduced. Hence, the RSGs themselves make excellent tools
with which to probe the global properties of their host clusters.

The clusters' distances from Earth and Galactic longitude place them
at Galacto-centric distances of $\sim$3.5\,kpc, and at an azimuthal
angle $\phi_{GC} = 45\pm10$\degr\ relative to the Sun-Galactic Centre
axis. This places them at the tip of the Galactic Bar as described by
\citet{Benjamin05}, which has half-length 4.5kpc and $\phi_{GC} =
44\pm10$\degr. This raises the intriguing possibility that the RSG
clusters were formed in a region-wide starburst event, triggered by
the interaction between the Bar and the disk. Aside from the two
clusters, there is certainly plenty of other evidence for recent
star-forming activity along this line-of-sight: D07 found several
other red stars in the field of RSGC2, which while not belonging to
the cluster, were likely to be supergiants and hence massive
stars. This is consistent with the large number of RSGs found at the
base of the Scutum spiral arm found by \citet{LC99}. Also noted in D07
were other objects indicative of recent star-forming activity, such as
the candidate supernova remnant IRAS~18369-0557 and the candidate
massive evolved star IRAS~18367-0556; while \citet{G-H08} have
identified the TeV gamma-ray source HESS~J1837-069 with possibly two
young pulsar wind-nebulae.

In this paper we assess the chemical abundances of the two clusters,
using high-resolution $H$-band spectroscopy of a sample of RSGs in
each cluster, in combination with LTE model atmospheres. Our method
allows us to derive abundances for C, Fe, and the $\alpha$-elements O,
Ca, Si, Mg, and Ti. We use these chemical abundances to address two
distinct topics.

Firstly, we use the relative surface abundances to test the
predictions of stellar evolutionary models. In massive stars, the
primary route of H fusion is via the CNO-cycle. Due to the relative
reaction rates involved in this cycle, a depletion of carbon occurs at
the expense of nitrogen. As stars enter the RSG phase, their deep
convective layers are expected to bring the products of nuclear
burning to the surface. Thus, the relative C abundances in the
atmospheres of the RSGs allow us to make critical tests of the latest
stellar evolution codes, such as those which include rotation
\citep[e.g.][]{Mey-Mae00} and the interaction with binary companions
\citep[e.g.][]{Eldridge08}.

Secondly, we will use the relative abundances of Fe and several
$\alpha$-elements in the RSG clusters, in combination with the
clusters' coincidence with the Galactic Bar, to investigate the
star-formation history of the Galaxy. The dominant source of $\alpha$
enrichment in the Interstellar Medium (ISM) is through type-II
supernovae (SNe), i.e.\ the core-collapse of massive stars. On the
other hand, Fe-peak elements are produced in type-Ia SNe --
thermonuclear explosions of low mass stars which reach the
Chandrasekhar limit through accretion from a companion. Hence,
$\alpha$-elements are enriched on short ($\sim$Myr) timescales, while
Fe-peak abundances are increased over much longer timescales
($\sim$Gyr).

In our Galaxy, we see variations in the [$\alpha$/Fe] ratio: in the
central Galactic Bulge, at Galacto-centric distances of $R_{\rm GC}
\la $2\,kpc, the [$\alpha$/Fe] ratio is found to be super-solar from
analyses of Red Giants
\citep[][]{R-O05,C-S06,Fulbright06,Fulbright07,Lecureur07}. In the
`thin' Galactic disk at $R_{\rm GC} \ga 5$\,kpc and scale height $|h|
\la $100pc, the ratio $\alpha$/Fe is found to be closer to solar
\citep[][ and references therein]{Bensby04,Luck06}. In addition, the
relative abundances in the thin disk are a function of $R_{\rm GC}$:
there is evidence to suggest heavy elemental abundance levels increase
at lower Galacto-centric distances, while the ratio [$\alpha$/Fe]
increases in the outer disk \citep[e.g][]{Luck06,Yong06}.

These abundance patterns are commonly explained as being due to the
distinct star-formation histories of the different Galactic
environments. The Bulge was formed rapidly several Gyrs ago, in a
starburst event which was too brief for the chemical evolution to be
affected by Type-Ia SNe. In the disk, star-formation has continued
over the lifetime of the Galaxy, allowing Type-Ia SNe to contribute to
the chemical enrichment of the ISM.

In the inner disk ($R_{\rm GC} \la 5$\,kpc), the abundance patterns
are thought to be strongly influenced by the Bar. External galaxies
with bars are found to have significantly flatter abundance gradients
than those without \citep{Mar-Roy94,Zaritsky94}. This is explained as
being due to radial motions induced by the bar potential
\citep{Friedli94,F-B95}. In addition, pile-up of material at the bar's
resonances can create star-forming hot-spots which result in localised
chemical enrichment. The RSG clusters themselves are evidence of such
a localised starburst in the inner regions of our own Galaxy, and have
Galactocentric distances similar to co-rotation ($\sim$3.5kpc) and the
Outer Lindblad Resonance ($\sim$5kpc).

In this paper, we will explore the chemical abundances of the inner
disk from analyses of the RSGs in the two Scutum clusters. By
comparing to other abundance measurements in the inner disk, we will
probe the star forming history in a region which is pivotal in the
Galaxy's secular evolution.

The paper is organized as follows: in Sect.\ \ref{sec:obs} we describe
our observations, data-reduction steps and analysis techniques; in
Sect.\ \ref{sec:res} we present our results and discuss the abundance
patterns of the two clusters. We make quantitative comparisons of
these results with stellar evolutionary models in Sect.\
\ref{sec:evol}, while in Sect.\ \ref{sec:gal} we discuss the
abundances of the two clusters in the context of those of the inner
Galaxy. We conclude in Sect.\ \ref{sec:conc}.

\section{Observations, data reduction \& analysis} \label{sec:obs}
Observations were taken during the night of 13th Aug 2006 using
NIRSPEC, the cross-dispersed echelle spectrograph mounted on Keck-II
\citep{McLean95}. The instrument was used in high-resolution mode with
the 0.576\arcsec\ $\times$ 12\arcsec\ slit and the NIRSPEC-5
filter. We used a dispersion angle of 63.0\degr\ and a
cross-dispersion angle of 36.72\degr. This achieved a spectral
resolution of $\sim$17,000 at select regions within the wavelength
range of 1.5-1.7\microns.

We used integration times of 20s, and observed each star twice at
different positions nodded along the slit. We observed all 14 RSGs in
RSGC1, and 12 RSGs from RSGC2. The identifications of the specific
stars observed, taken from \citet{Figer06} and \citet{RSGC2paper}, can
be found in Table \ref{tab:data}. In addition to the cluster stars, we
also observed HD 171305 (spectral type B0~V) as a telluric
standard. Flat-field images were taken with a continuum lamp, while
for wavelength calibration we observed Ar, Ne, Xe and Kr arc lamps, to
provide a large number of template lines in the narrow wavelength
range of each spectral order.

Removal of sky emission, dark current and bias offset was done by
subtracting nod-pairs, and images were flat-fielded with the
continuum-lamp exposures. The spectral traces produced by NIRSPEC are
warped in both the spatial and dispersion directions, so before
extracting the spectra from the reduced science frames each
two-dimensional order was rectified onto a linear
grid. The spatial warping was characterized by the spectral traces of
the two stellar traces in a nod pair. The distortion in the dispersion
direction was fitted using the arc- and etalon-lines. For technical
details of the rectification process, see \citet{Figer03}.

As we know the wavelengths of the arc-lines, rectification also
wavelength calibrates the data. By rectifying the arc frames and
measuring the residuals of the arc-line wavelengths, we found the
wavelength solution to be accurate to $\pm$4\kms\ across all orders. 

Spectra were extracted by summing each channel in the rectified orders
within $\pm$5$\sigma$ of the centre of each spectrum. Cosmic-ray hits
and bad pixels were identified by comparing the two nod spectra of
each star, and were replaced by the median value of the closest 4
pixels. 

We removed the H and He~{\sc i} absorption features of the telluric
standard via linear interpolation either side of the line. The
atmospheric absorption features in the science frames were then
removed by dividing through by the telluric standard. Finally, the
spectra were normalised by dividing through by the median continuum
value. From featureless continuum regions in the final spectra, we
estimate the signal-to-noise to be better than 100 for all spectra.

\subsection{Data analysis} \label{sec:anal}
In order to obtain chemical abundances from the spectra, we have
developed a technique of generating synthetic spectra from cool star
model atmospheres, and adjusting the input relative abundances to
match the equivalent widths of certain diagnostic lines. The code used
is that presented in \citet{Origlia93}, which was updated in
\citet{Origlia02,Origlia03}. The code has been used successfully to
obtain abundances of Bulge giants \citep[e.g.][]{R-O05}, young
clusters dominated by RSGs \citep[e.g.][]{Larsen08}, and most recently
to study the chemical abundandes of RSGs in the Galactic Centre
\citep{RSGGC}.

The code and methodology is explained in detail in the papers listed
above. Briefly, the code uses the LTE approximation and model
atmospheres from the grid of \citet{Johnson80} which incorporate the
effect of molecular blanketing. Thousands of near-IR molecular
ro-vibrational and atomic transitions are included. Initial estimates
of the stars' luminosities, temperatures and surface gravities were
estimated based on the results of D07 and D08, while the
microturbulent velocity $\xi$ is estimated from the OH and CO
lines. We were able to produce satisfactory fits to the data without
the inclusion of macroturbulence. This is to be expected for
observations of RSGs with initial masses $\sim$10-20\msun, whose
macroturbulent velocities are often found to be comparable to the
spectral resolution of our observations \citep[see also][]{RSGGC}.

Abundance estimates are obtained by adjusting relative abundances in
the model atmospheres, computing the synthetic spectra, and minimizing
the residuals between the observed and synthetic spectra in the
equivalent widths of certain diagnostic lines. We concentrate on
fitting those lines which are relatively unblended and for which
reliable atomic/molecular data exist. The ro-vibrational lines of OH
and CO are used to determine oxygen and carbon abundances, while
abundances of other metals are derived from the atomic lines of Fe~I,
Si~I, Mg~I, Ca~I, and Ti~I. While the molecule CN is included in the
analysis, severe blending of the lines at the observed spectral
resolution prevents an accurate determination of the N abundance.  We
do however find results that are consistent with CNO equilibrium, that
is with N being enhanced by the same level that C is depleted. For
compilations of atomic/molecular data we use the Kurucz
database\footnote{\tt
  http://cfa-www.harvard.edu/amdata/ampdata/kurucz23/sekur.html},
\citet{B-G73}, \citet{H-H79}, \citet{S-L82}, and \citet{M-B99}.

\begin{figure*}
  \centering
  \includegraphics[width=12cm,bb=58 144 572 698]{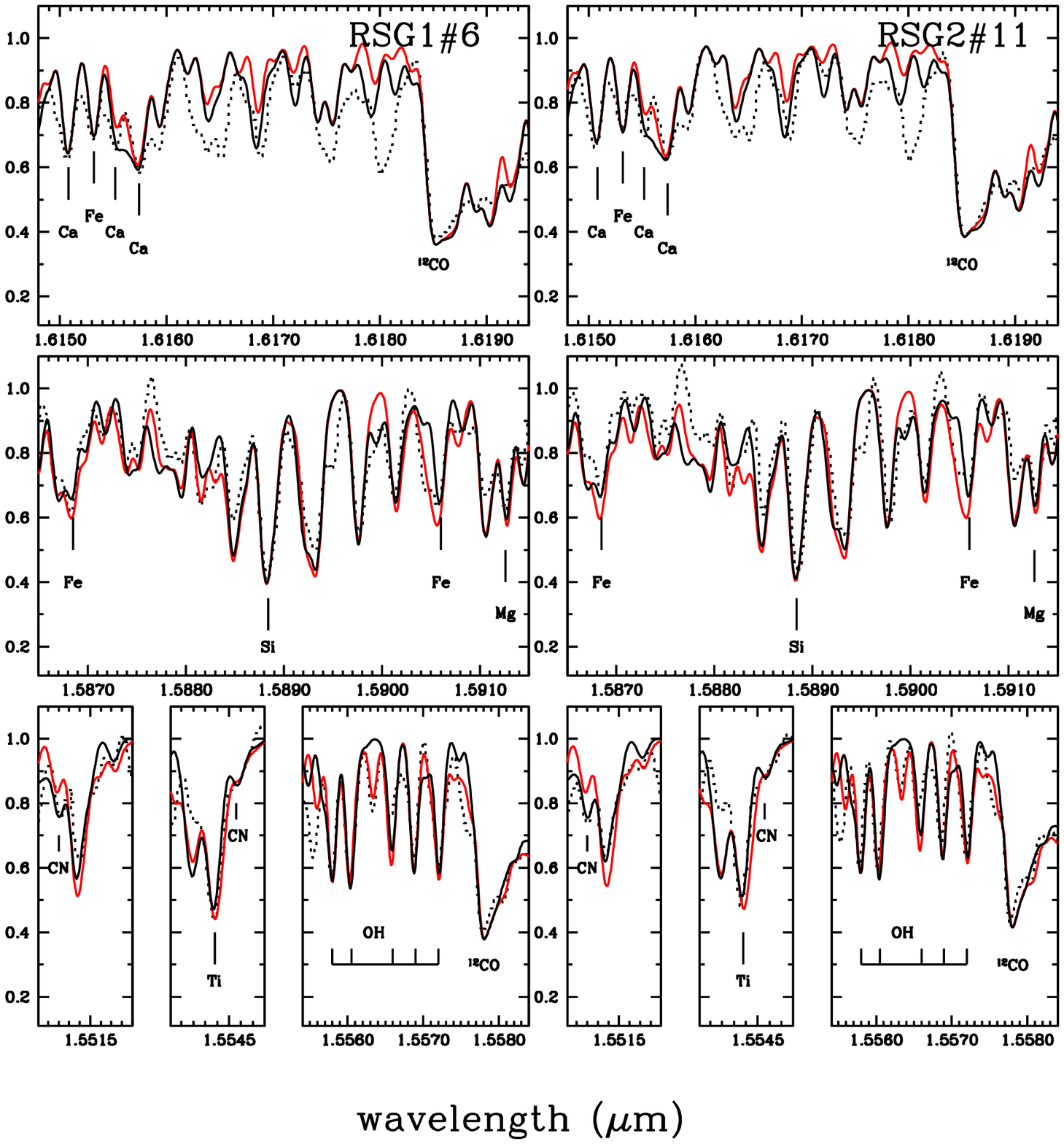}
  \caption{Examples of the H-band spectra of one star in each cluster,
    showing select regions and the identified lines. Observations are
    shown as dotted lines, our best-fit synthetic spectra are shown as
    solid black lines. In addition, we also plot the computed best-fit
    spectrum using the Uppsala CN line-list (B.~Gustavsson,
    priv.~comm. -- solid red lines), with negligible differences in
    the strengths of the diagnostic lines. }
  \label{fig:specfit}
\end{figure*}

For each best-fitting model, we estimate the uncertainties and
solution-robustness by generating a further four models with
abundances $\pm$0.1dex and $\pm$0.2dex. In addition, test models are
generated in which the stellar parameters are varied and the
abundances re-tuned, to assess the degeneracy of the model
solutions. In studying the residuals between the test-models and the
observed spectra, we find that solutions of similar statistical
significance are found with $\Delta$T$_{\rm eff} \pm$200\,K, $\Delta
\log g = \pm 0.3$, and $\Delta \xi = \mp$0.5\kms, which we take to be
the uncertainties on the stellar parameters. We find that the measured
abundances are stable across the test models to within 0.1-0.2dex,
while we found that gravities $\log g = 0.0 \pm 0.3$ and
microturbulence $\xi = 3 \pm 0.5$\kms\ produced satisfactory fits for
all stars modelled. The quoted errors include the uncertainty on the
continuum placement when measuring the line equivalent widths, which
can be problematic in H-band spectra of cool stars due to the number
of spectral features. As a further cross-check on our results, we
re-ran the spectral sysnthesis using the CN line-list provided by
B.~Gustafsson (priv.~comm.), finding negligible differences in derived
abundances (see Fig.\ \ref{fig:specfit}). 

Finally, as further validation of our method we note that in our
recent study of two RSGs in the Galactic Centre, we found that our
results were consistent with those of complimentary studies of the
same objects to within our quoted uncertainties \citep{RSGGC}, while
in a study of Arcturus we found abundances consistent with previous
measurements \citep{R-O05}.

\section{Results} \label{sec:res}
Selected regions of the H-band spectra of one star in each cluster are
shown in Fig.\ \ref{fig:specfit}. Each show a dense absorption line
spectrum, dominated by neutral lines of Fe, several $\alpha$ elements
(Si, Ca, Mg, Ti), as well as molecular bands of OH and CO. Below, we
describe the abundance patterns of the elements studied. As reference,
we use the solar abundances of \citet{Asplund05}.

The elemental abundances we derive for each star are listed in full in
Appendix A. To within the uncertainties, the elemental abundances for
each star studied in each cluster are consistent. In Table
\ref{tab:mean}, we show the average abundance for each element in each
cluster. Here, it can be seen that the average abundances in the two
clusters are similar to within the errors. This evidence, together
with the proximity of the clusters to one another ($d_{sep} =
0.8^{+1.6}_{-0.7}$kpc, D08) and their similar ages (D07, D08),
suggests that the two clusters formed out of the same giant molecular
cloud.

\begin{deluxetable*}{lccccc}
\centering
\tabletypesize{\scriptsize} \tablecaption{Average elemental abundances
of the two clusters, where $A(X) = \log(X/H)+12$. Columns show (1) the
elements studied; (2) the mean solar abundances, taken from
\citet{Asplund05}; (3) \& (4) the average abundances for RSGC1 and
RSGC2 respectively; and (5) the ratio of each element's average
abundance from the two clusters compared to Fe, normalized to the
solar value. The uncertainties in the average cluster abundances are
taken as the rms standard deviation in each value. \label{tab:mean}}
\tablewidth{16cm} \tablehead{ \colhead{Element $X$}&
\colhead{$A(X)_{\odot}$}& \colhead{$A(X)_{\rm RSGC1}$}&
\colhead{$A(X)_{\rm RSGC2}$}& \colhead{[$X$/Fe]$_{\rm RSGC1}$}&
\colhead{[$X$/Fe]$_{\rm RSGC2}$} } \startdata 
Fe &       7.45 &       7.33 &       7.28 & -- & -- \\
-- & $\pm$ 0.05 & $\pm$ 0.05 & $\pm$ 0.03 & -- & -- \\
 O &       8.66 &       8.61 &       8.59 &       0.08 &       0.10 \\
-- & $\pm$ 0.05 & $\pm$ 0.07 & $\pm$ 0.05 & $\pm$ 0.11 & $\pm$ 0.09 \\
Si &       7.51 &       7.30 &       7.36 &      -0.08 &       0.02 \\
-- & $\pm$ 0.04 & $\pm$ 0.08 & $\pm$ 0.08 & $\pm$ 0.12 & $\pm$ 0.11 \\
Mg &       7.53 &       7.41 &       7.36 &       0.00 &      -0.00 \\
-- & $\pm$ 0.09 & $\pm$ 0.13 & $\pm$ 0.09 & $\pm$ 0.17 & $\pm$ 0.14 \\
Ca &       6.31 &       6.20 &       6.18 &       0.02 &       0.04 \\
-- & $\pm$ 0.04 & $\pm$ 0.05 & $\pm$ 0.04 & $\pm$ 0.10 & $\pm$ 0.08 \\
Ti &       4.90 &       4.92 &       4.92 &       0.14 &       0.19 \\
-- & $\pm$ 0.06 & $\pm$ 0.10 & $\pm$ 0.06 & $\pm$ 0.14 & $\pm$ 0.08 \\
 C &       8.39 &       7.95 &       7.89 &      -0.32 &      -0.33 \\
-- & $\pm$ 0.05 & $\pm$ 0.07 & $\pm$ 0.06 & $\pm$ 0.11 & $\pm$ 0.10 \\
\enddata
\end{deluxetable*}

We find that, in general, both clusters are slightly metal-poor with
respect to the solar value. The Fe content averaged over the two
clusters is $\rm [Fe/H] = -0.15 \pm 0.07$. The $\alpha$-elements (Ca,
O, Mg, Si, Ti) also show abundances which are $\sim$0.1-0.2dex below
the solar level. These values do not take into account self-depletion
of H at the surfaces of the stars: in \citet{RSGGC} we show, with the
aid of evolutionary models, that the atmospheres of RSGs can be
depleted in H by up to 0.1dex, as the deep convective layers draw the
nuclear-processed material to the surface. Therefore, we conclude that
the {\it initial} [Fe/H] and [$\alpha$/H] ratios of the two clusters
were $\approx$0.2-0.3dex below solar.

The ratio of the $\alpha$-elements to Fe, which is a powerful probe of
star-formation history (see Sect.\ \ref{sec:intro}), is consistent
with solar for each of Si, Mg, Ca and Ti. We find $\rm [O/Fe] = +0.09
\pm 0.11$ using the \citet{Asplund05} solar abundances. When the
\citet{G-S98} solar values are used, we find $\rm [O/Fe] = -0.03 \pm
0.11$. Given the recent fluctuations in the derived solar O abundance,
at this time we conclude for now that the O/Fe ratio in the clusters
is consistent with solar.

\begin{figure*}
  \centering
  \includegraphics[width=8cm,bb=0 20 546 453]{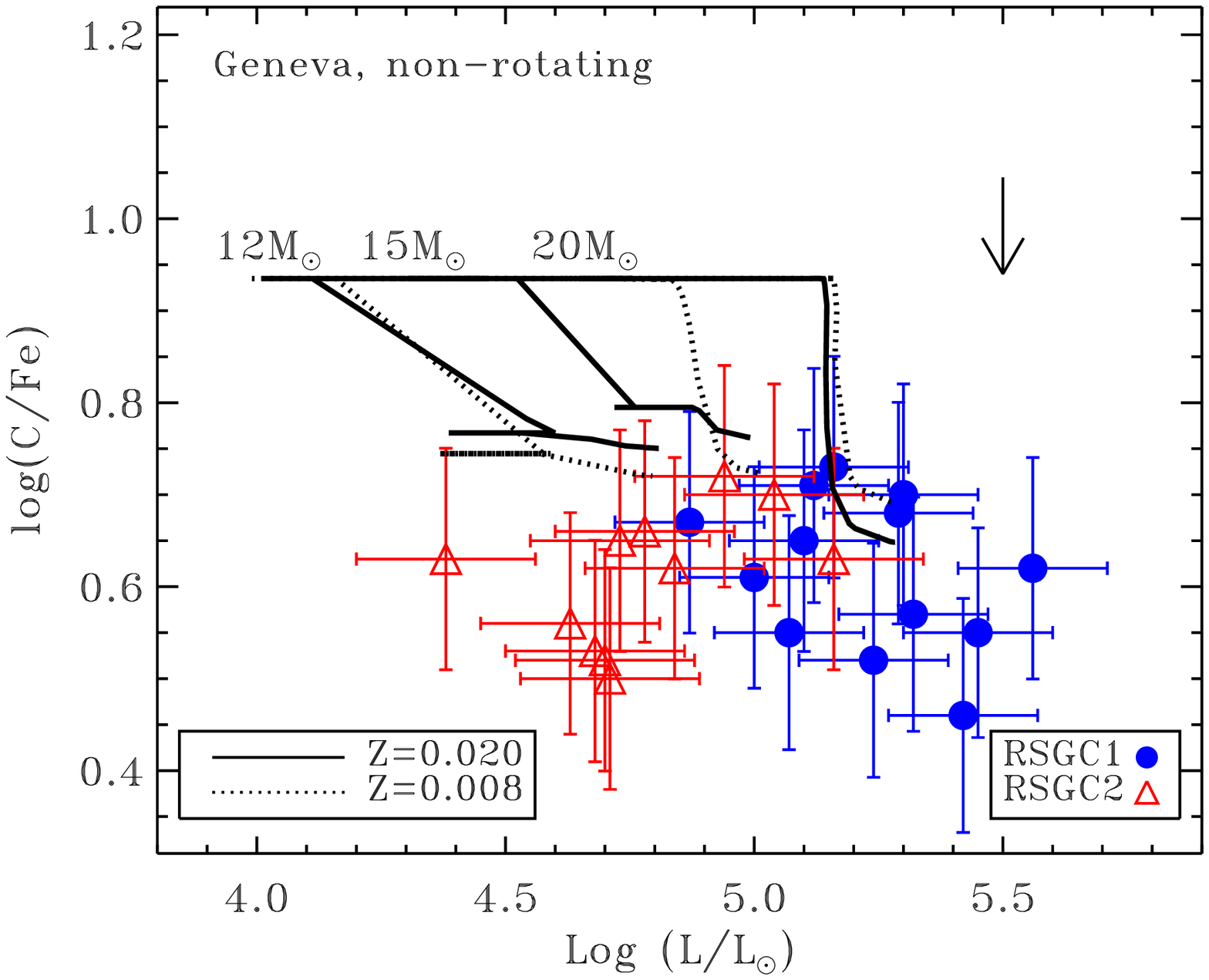}
  \includegraphics[width=8cm,bb=0 20 546 453]{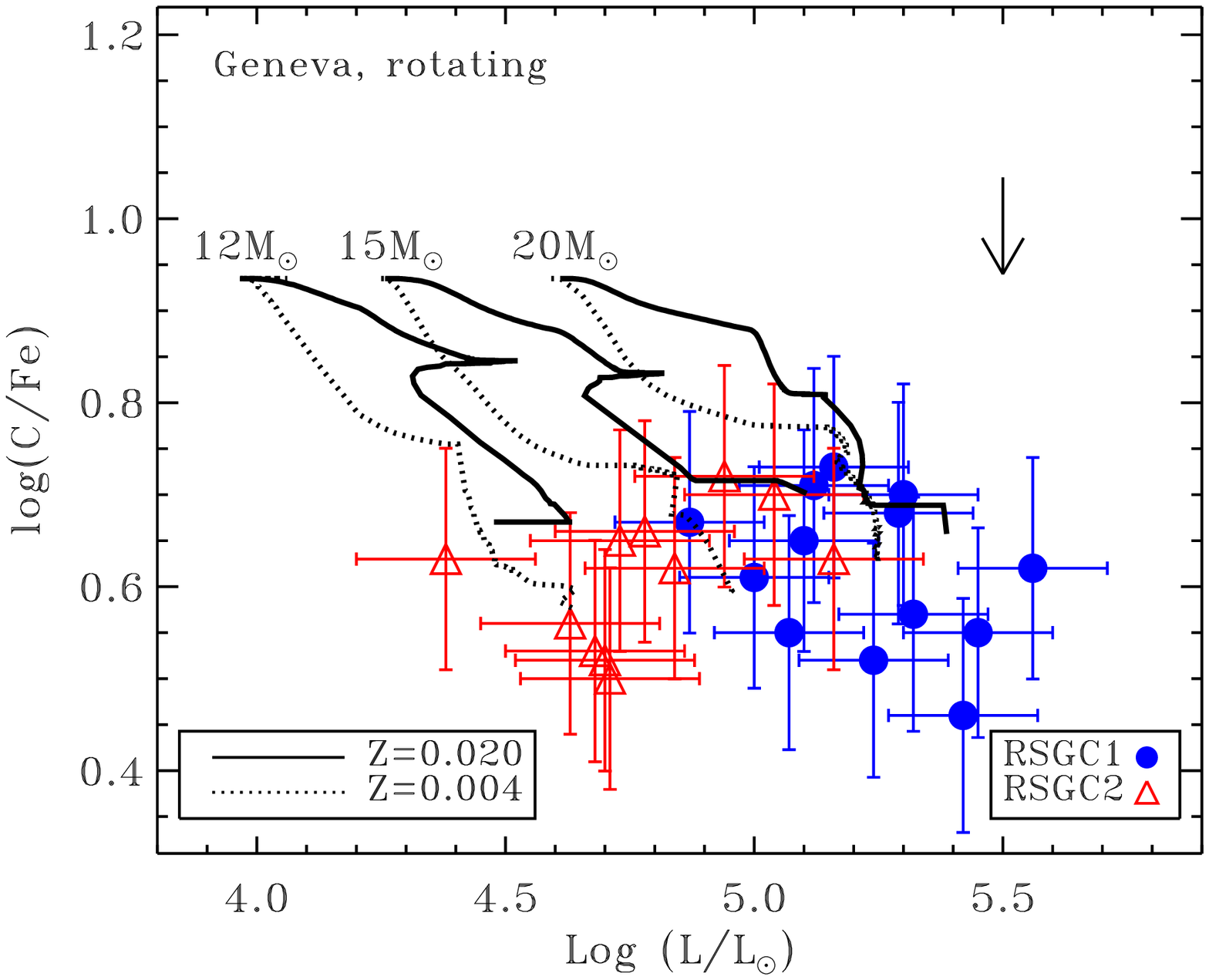}
  \caption{Carbon fraction as a function of stellar
  luminosity. Overplotted are mass-tracks from the Geneva non-rotating
  ({\it left}) and rotating ({\it right}) isochrones at solar and
  sub-solar metallicities (see text for references). Note that the
  sub-solar metallicities are different in each panel. The mass-tracks
  have been shifted downwards to reflect the recent revision in the
  Solar [C/Fe] -- the arrow in the top-right of each panel indicates
  the magnitude of this change (see text for details). \\ }
  \label{fig:clumrot}
\end{figure*}

The levels of carbon in the stars' atmospheres appear to be
significantly depleted, consistent with their evolved status. This
result is expanded upon in detail in Sect.\ \ref{sec:evol}.

\section{Comparison with stellar evolutionary models} \label{sec:evol}
The results presented in Sect.\ \ref{sec:res} allow us to test various
predictions of stellar evolutionary models. As stars enter the RSG
phase, the products of CNO burning are expected to be seen at the
surface, specifically the enrichment of N at the expense of C. From
the values in Table \ref{tab:data}, it can be seen that all stars in
both clusters show evidence of C depletion. In Fig.\ \ref{fig:clumrot}
we investigate quantitatively the predictions for surface C abundances
as a function of evolutionary state.

Throughout our analysis, we consider the C fraction in relation to the
Fe abundance. In evolutionary models, the relative abundance of an
element $X$ with respect to hydrogen $X$/H depends on two input
parameters: the relative abundances of the heavy elements, and the
ratio of the heavy element abundances to H (i.e.\ the metallicity,
$Z$). However, the ratio $X$/Fe depends only on the relative
abundances of the heavy elements. Thus, by comparing C as a function
of Fe, we reduce the model-dependency of our analysis. As the relative
abundances of the heavier elements in the stars studied here appear to
be consistent with solar (see Table \ref{tab:mean}), we use models
with solar-scaled heavy element abundances.

\subsection{Non-rotating models vs rotating models}
In Fig.\ \ref{fig:clumrot} ({\it left}) we plot the C/Fe fraction as a
function of luminosity for the stars observed. The stellar
luminosities are taken from D07 and D08. Overplotted are mass tracks
at initial masses of 12\msun, 15\msun\ and 20\msun\ from the
non-rotating Geneva models. For reference, the initial masses of the
RSGs in each cluster (derived from the rotating Geneva models) are
18\msun\ for RSGC1 and 14\msun\ for RSGC2 (D08). As we find a
sub-solar iron content for the stars in each cluster, we plot tracks
using both solar metallicity \citep[Z=0.02 ][]{Schaller92} and Z=0.008
\citep[][]{Schaerer93}. The iron content we derive for the RSG
clusters indicates a metallicity somewhere between these two
values. In the right-hand panel of Fig.\ \ref{fig:clumrot}, we plot
the same quantities in comparison to the Geneva models which include
rotation, at $Z=0.02$ and $Z=0.004$ \citep[][ respectively]{Mey-Mae00,
M-M01}\footnote{Geneva rotating models at Z=0.008 are unavailable at
the time of writing.}

When comparing our derived C abundances with those predicted by the
Geneva evolutionary models, it is important to first note that these
models use the relative heavy elemental abundances of
\citet{G-S98}. In the 3-D solar model of \citet{Asplund05} the
fractional abundances of C and Fe have been revised slightly. For this
analysis we assume that this small change in relative metal abundances
would have no major impact on the output of the Geneva evolutionary
code other than to cause a linear displacement in abundance-space. We
have shifted each mass track by 0.105dex -- the magnitude of the
revision in the Solar [C/Fe] fraction. We indicate the magnitude of
this displacement with an arrow in the upper-right of each of the
panels in Fig.\ \ref{fig:clumrot}.

In general, the non-rotating models do not quite reproduce the
observed level of C depletion in the RSGs of the two Scutum clusters,
even when the updated solar [C/Fe] is taken into account
(Fig.\ \ref{fig:clumrot}, {\it left}). There seems to be much better
agreement between the data and the rotating models, which are more
heavily depleted than the non-rotating models due to
rotationally-enhanced internal mixing. Sub-Solar metallicity models
suffer a slightly greater level of C depletion, due to decreased
mass-loss on the main sequence which inhibits the loss of angular
momentum and increases the level of internal mixing. This can be seen
in the right-hand panel of Fig.\ \ref{fig:clumrot}, where the low-Z
mass-tracks have systematically lower C/Fe ratios. While we are unable
to discern between the Solar and sub-Solar mass-tracks, we can say
that the rotating models provide a much better fit to the [C/Fe]
levels, while it has already been shown that these models do a better
job of reproducing the luminosity distributions of the two clusters
(D08).

\subsection{Binary models}
To assess the impact of binarity on surface abundances we refer to the
Cambridge binary star evolution code presented in
\citet{Eldridge08}. The code includes the effect of Roche Lobe
Overflow and Common-Envelope Evolution, processes which are
particularly important when the primary star evolves to the RSG phase
due to the expanded atmosphere.

Briefly, the evolution to the RSG phase of a massive star can result
in the star filling its Roche Lobe and transferring matter to the
companion star. This reveals the lower layers of the primary's
atmosphere in which the effects of CNO burning are more
evident. However, mass-transfer of the primary's envelope onto the
secondary also restricts the primary's redward evolution, resulting in
drastically shortened RSG lifetimes. \citet{Eldridge08} showed that,
for a sample of binary systems with broad distributions of separations
and mass-ratios, the {\it mean} RSG lifetime averaged across the
sample was reduced by a factor of $\sim$3. However, for low-separation
binary systems the RSG lifetime-reduction is clearly much more
severe. Therefore, the chances of observing a RSG in a close binary
are extremely small, given the brevity of the phase in single stars
($t_{\rm RSG} \sim$1Myr for a single star in the Eldridge et al.\
models).

While binary models are able to increase the level of C depletion, the
drastically reduced RSG lifetimes lead us to conclude that close
binary interaction cannot have played a significant role in the
evolution of the RSGs in the two clusters. This is not to say that
binarity is not prevalent in the clusters, more that the RSGs in the
two clusters that we see today are unlikely to have evolved in a close
binary system. This has two important implications, which we now
discuss.

Firstly, in F06, D07 and D08, the masses of the clusters were
estimated by counting the numbers of RSGs, comparing to population
synthesis models created using single star evolutionary codes, and
extrapolating over the rest of the Initial Mass Function. If the
clusters have high binary fractions, then a significant number of
cluster members which had the same initial masses of the RSGs we see
now will have skipped the RSG phase. The not-unreasonable parameters
for a binary distribution used by Eldridge et al.\ suggest a reduction
in the number of RSGs by a factor of $\sim$3 compared with single-star
models. Therefore, the total cluster masses may have been underestimed
by a similar factor, implying initial masses in excess of $10^5$\msun,
which would make them the most massive young clusters in the Galaxy.

Secondly, the mere fact that we are looking at RSGs in these clusters
suggests that we are dealing with stars which have in effect evolved
in isolation. That is, it is reasonably safe to assume that the stars
are {\it single}. This is an extremely important consideration if we
are to use the stars to measure the velocity dispersion of the
clusters, and hence determine their virial masses. Contamination by
the high orbital velocities in close massive binaries would lead to a
significant systematic overestimate of a sample's velocity dispersion,
and hence of the virial mass. On the one hand, this should serve as a
note of caution when interpreting the velocity dispersions in young
massive clusters when measured from blue stars, particularly as the
massive binary fraction of Westerlund~1 has recently been inferred to
be very high \citep{Clark08}. On the other hand, this would also seem
to suggest that RSG spectral features provide the most reliable
measurement of cluster velocity dispersions \citep[see
  e.g.][]{Mengel08}.

\section{The RSG clusters and the star-formation history of the
  Galaxy} \label{sec:gal} 

In this section, we compare the relative abundances derived for the
RSG clusters with the standard indicators of star-formation history,
namely the ratios [Fe/H] and [$\alpha$/Fe]. In broad terms, the
primary source of $\alpha$-element enrichment in the Galaxy is through
Type-II supernovae (SNe), that is the core-collapse of stars with
masses greater than $\sim$8\msun. On the other hand, the main source
of Fe enrichment is SNe Ia -- thermonuclear explosions of low mass
stars which reach the Chandrasekhar limit through accretion from a
companion. Hence, as Fe- and $\alpha$-enrichment occurs on vastly
different timescales, star-formation histories can be probed through
comparisons of the $\alpha$ and Fe abundances

\subsection{The abundance patterns of the Galaxy}
In the Galactic `thin' disk, with Galacto-centric distances $R_{GC}
\ga$4kpc and scale height $ |h| \la$ 100pc, it has been found from
abundance analysis of Cepheids and late-type dwarfs that the ratio
[$\alpha$/Fe] is approximately solar \citep[][and references
  therein]{Bensby04,Luck06}. Meanwhile, in the `thick' disk ($ |h|
\ga$ 100pc), halo and bulge, [$\alpha$/Fe] is found to be super-solar
from analyses of Red Giants
\citep[][]{R-O05,C-S06,Fulbright06,Fulbright07,Lecureur07,Melendez08}.

In addition, many authors have found evidence for a trend of
increasing chemical abundances with decreasing $R_{GC}$, the so-called
Galactic metallicity gradient. Analyses of B dwarfs
\citep{Rolleston00,Smartt01}, Cepheids \citep{Luck06}, UC-HII regions
\citep[e.g.][]{Afflerbach97}, and Planetary Nebulae
\citep[e.g.][]{M-Q99} have all found evidence for increasing metal
abundances, both of Fe and $\alpha$-elements, at lower
$R_{GC}$. Further, the results of Andrievsky and collaborators
\citep[summarised in ][]{Luck06}, who studied the abundances of both
Fe and the $\alpha$-element Ca in Cepheids, showed that [Ca/Fe] was
solar and roughly constant throughout the thin disk out to $R_{\rm GC}
\approx 8$kpc, with tentative evidence for an increase in [Ca/Fe]
in the outer Galaxy.

These abundance patterns are thought to be a result of the Galaxy's
evolution and star-formation history: the bulge and halo formed very
quickly \citep[$\sim$0.1-0.5Gyr,][]{Ballero07}, during which time the
chemical evolution was dominated by SNe II. After this time, star
formation ceased and the $\alpha$-heavy abundances were
frozen-in. However, in the Galactic disk star-formation continued at
some rate which was a function of $R_{GC}$. Eventually the enrichment
of the Fe abundance by SNe~Ia brought the $\alpha$/Fe ratio to below
that of the halo and bulge.

\begin{figure}[t]
  \centering
  \includegraphics[width=8.5cm,bb=0 0 623 400]{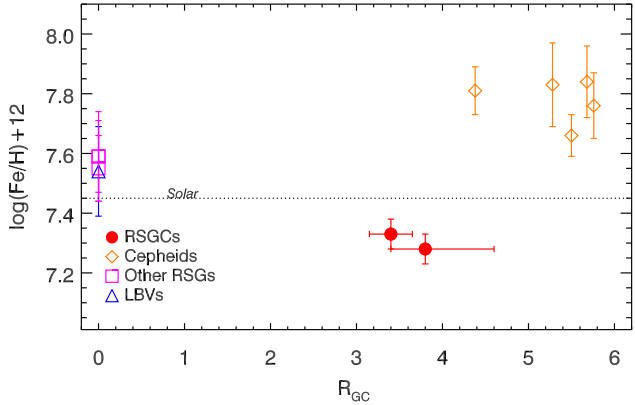}
  \caption{The trend of Fe/H abundance in the inner 6kpc of the
  Galaxy. The dotted line in each panel indicates the relevant solar
  value from \citet{Asplund05}. The references for the literature data
  are given in Table \ref{tab:refs}. }
  \label{fig:fegrad}
\end{figure}

\begin{figure}[t]
  \centering
  \includegraphics[width=8.5cm]{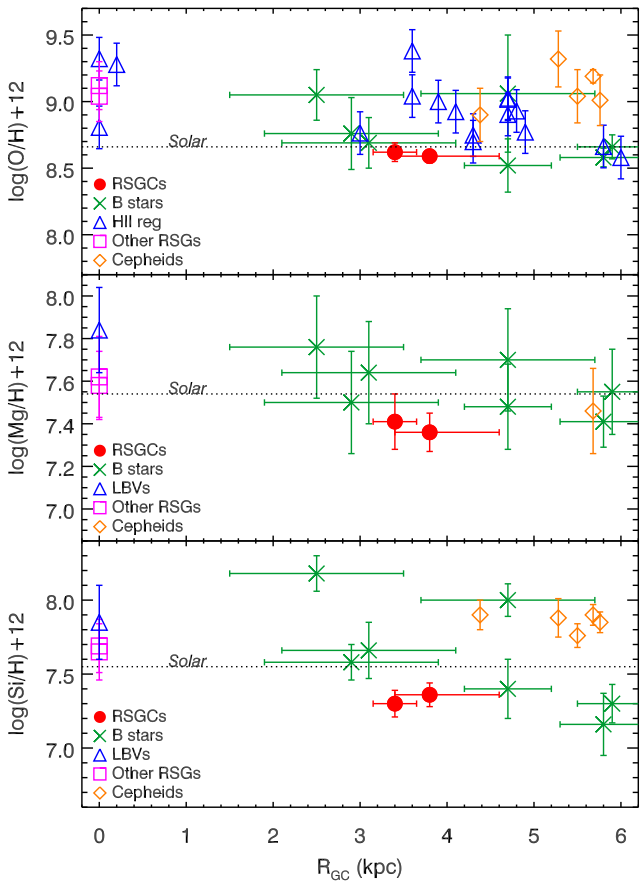}
  \caption{Same as Fig.\ \ref{fig:fegrad} but for oxygen (top panel),
  magnesium (middle panel) and silicon (bottom panel). }
  \label{fig:zgrad}
\end{figure}

\begin{deluxetable}{ll}
\tabletypesize{\scriptsize} \tablecaption{References for the data
  shown in Figures \ref{fig:fegrad} - \ref{fig:azgrad}. \label{tab:refs}}
\tablewidth{8cm}
\tablehead{
\colhead{Object}&
\colhead{References}
}
\startdata
B stars            &  \citet{Smartt01,D-C04} \\
H{\sc ii} regions  &  \citet{Afflerbach97}   \\
Other RSGs         &  \citet{RSGGC,Cunha07}  \\
LBVs               &  \citet{Najarro08}      \\
Cepheids           &  \citet{Andrievsky02}  \\
\enddata
\end{deluxetable}

\subsection{Comparison between Galactic and RSGC abundances}

\subsubsection{Iron} \label{sec:iron}
As stated in Sect.\ \ref{sec:res}, we find that the average Fe/H
abundance ratios for the two clusters are 0.1-0.2dex below solar. This
is neglecting the effects of self-depletion in the atmopsheres in
RSGs, which we estimate to be a further $\sim-0.05-0.1$dex for stars
with initial masses of 14-18\msun\ \citep[see Fig.\ 2 of ][]{RSGGC}.

In Fig.\ \ref{fig:fegrad} we compare the Fe/H ratios of the clusters
with other Fe/H measurements in the inner Galaxy. In analysis of the
atmospheres of Cepheids, \citet{Andrievsky02,Luck06} find that the
Fe/H ratio increases towards the Galactic Centre (GC) at a rate of
-0.06dex\,kpc$^{-1}$, reaching $\approx$+0.2dex above solar at $R_{\rm
  GC} \la 6$kpc. In the GC, recent studies suggest that the Fe
abundance is roughly solar \citep[e.g.]{Najarro08,RSGGC}. We note that
\citet{Cunha07} found slightly super-solar Fe levels from studies of
RSGs in the GC. However, the initial Fe/H ratios of Cunha et al.'s
sample were likely 0.05-0.1dex lower than reported due to the
affore-mentioned self-depletion effect. 

From Fig.\ \ref{fig:fegrad} it is clear that the RSGCs do not follow
the radial Galactic Fe/H trend -- the clusters exhibit Fe abundances
which are $\sim$0.5dex below those of the Cepheids in the inner
disk. This plot may simply illustrate the inadequacies of a
one-dimensional parameterization of abudance trends, particularly in
the inner Galaxy. The presence of physical structures in the inner
regions of the disk, such as the Bar, and the ends of the spiral arms,
represent large azimuthal fluctuations in stellar/gas density and
star-formation rates. As such, it seems unlikely that chemical
abundances should be homogeneous around the inner disk. This is a
point that we investigate further in Sect.\ \ref{sec:azvar}.

\subsubsection{The $\alpha$-elements: O, Ca, Si, Mg, and Ti }
As with Fe, these elements are again found to be sub-solar in relation
to H. However, the ratio of $\alpha$/Fe is solar to within the errors,
consistent with a disk population. In Fig.\ \ref{fig:zgrad} we again
put these results in context with measurements of objects at similar
$R_{\rm GC}$. In a search of the literature, we have found abundance
measurements in the inner disk for O, Mg and Si for Cepheids
\citep{Andrievsky02}, B~dwarfs \citep{Smartt01,D-C04}; while we also
find O measurements from H~{\sc ii} regions \citep{Afflerbach97}. The
H~{\sc ii} region measurements are less reliable as they are highly
model-dependent -- they determine temperatures by matching the amount
of ionizing UV flux (inferred from the radio) to model stellar
atmospheres, which in itself depends on metallicity. Nevertheless, we
include their data for completeness.

\begin{figure}[t]
  \centering
  \includegraphics[width=8.5cm,bb=0 0 480 340]{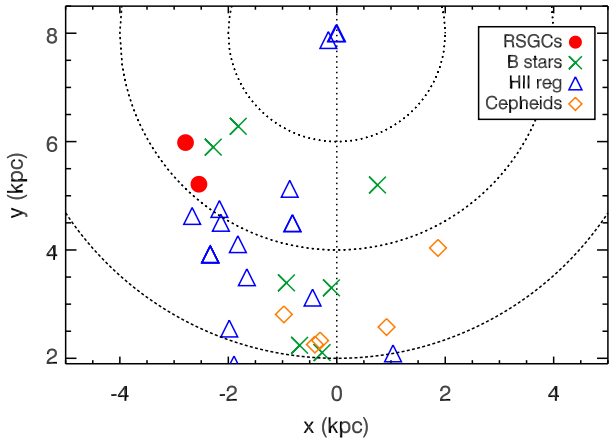}
  \caption{Location in the Galaxy of the B~stars, \hii~regions
  and Cepheids used as probes of the metallicity gradient. We assume
  the Galactic centre to be 8\,kpc from Earth. The dotted circles show
  Galacto-centric distances of 2, 4, and 6\,kpc. The references for
  the literature data are given in Table \ref{tab:refs} }
  \label{fig:radplot}
\end{figure}

Figure \ref{fig:zgrad} again shows that the RSG clusters appear to
have systematically lower abundances than other objects located at
similar Galacto-centric distances. However, when the B~dwarf
measurements are included, the discrepancy of the RSGC data appears
less striking. Indeed, taken together, the cepheid, RSGC and B~dwarf
measurements seem to suggest that there be large abundance variations
in the inner Galaxy. 

\subsubsection{Azimuthal variations in relative
  abundances} \label{sec:azvar} 
As stated in the preceeding sections, the simple one-dimensional
parameterization of abundance patterns in a metallicity gradient may
be too obtuse when dealing with the inner regions of the
Galaxy. Indeed, the presence of the Bar at $R_{\rm GC} \la 4$kpc
implies large azimuthal fluctuations in stellar/gas density, hence it
is not inconceivable that variations in abundances may also exist.

\begin{figure}[t]
  \centering
  \includegraphics[width=8.5cm]{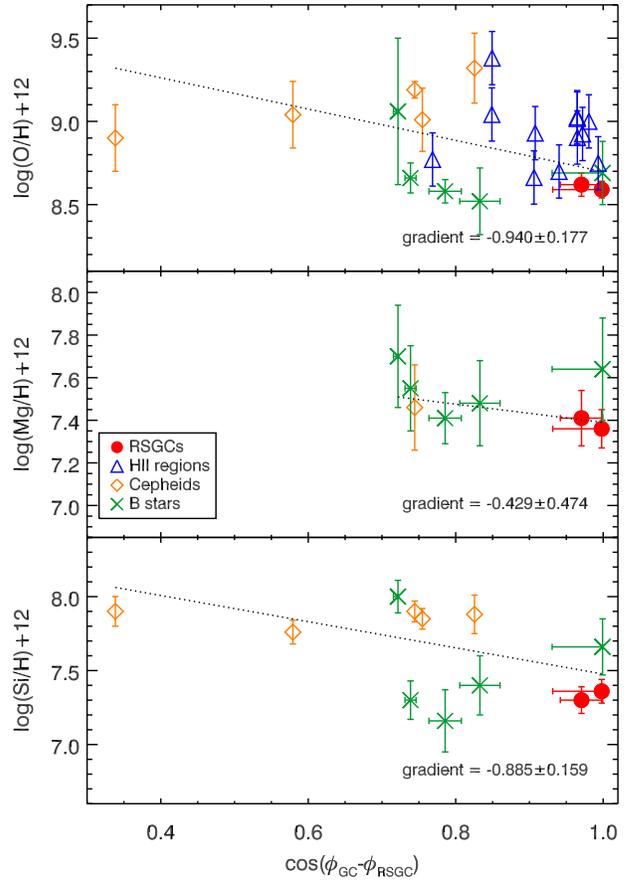}
  \caption{Azimuthal abundance gradients for oxygen, magnesium and
    silicon, created using objects with Galacto-centric distances of
    3.5 to 6.0\,kpc. The coordinate system has been rotated such that
    $(\phi_{\rm GC}-\phi_{\rm RSGC}) \equiv 0$\degr. The references for
    the literature data are given in Table \ref{tab:refs} }
  \label{fig:azgrad}
\end{figure}

In Fig.\ \ref{fig:radplot}, we show the locations of the objects
plotted in Figs.\ \ref{fig:fegrad} and \ref{fig:zgrad} in the X-Y
plane. We see that the Cepheids and the RSGCs, which have abundance
differences of $\sim$0.4dex, are separated in physical space by
approximately 4\,kpc. Curiously, the B~dwarfs and H~{\sc ii} regions,
which have intermediate chemical abundances between the RSGCs and
Cepheids, are also located between the two in physical space. This
raises the intriguing possibility of large-scale ($\sim$kpc) azimuthal
abundance variations in the inner disk.

To explore this prospect further, in Fig.\ \ref{fig:azgrad} we plot
the relative abundances of the Cepheids, B~dwarfs, \hii\ regions and
RSGCs as a function of their azimuthal angle about the Galactic
Centre, \phigc. We choose to plot the abundances as a function of the
cosine of the angle, as any periodic variations should show up as
linear trends in $\cos$(\phigc)-space. We define the positive
direction as being clockwise when viewed from above, and we have
rotated the coordinate system such that the RSG clusters, which have
the lowest abundances, have $\cos(\phi')=1$, where $\phi' \equiv
\phi_{\rm GC}-\phi_{\rm RSGC}$. In practice this coordinate system
transformation made very little difference to the results.

From Fig.\ \ref{fig:azgrad} we see that, though there is no strong
evidence of a gradient in Mg, trends do seem to exist in O and
Si. This is true even if the model-dependent \hii-region points are
removed. We find that the gradients in O and Si, determined from
linear regressions, are significant at the $\sim$5$\sigma$
level. Further, the gradients are similar in each element to within
the uncertainties. When the coordinate system was {\it not} rotated
(see above), this gradient significance was still in excess of
4$\sigma$.

These measurements are probing a limited range in azimuth, so these
observed gradients are representative of localised fluctuations in
abundance levels in the inner disk. Moreover, these fluctuations
apparently coincide with the end of the Bar. Given that we find these
statistically-significant variations, we identify three statistical
explanations for the observed trends which we discuss below.

\paragraph{Artifacts of random sampling --} If the abundances in the inner
Galaxy are uniform, but have large scatter, then a gradient in
abundances may be artificially inferred if we happen to sample the
extremes of the abundance distribution at the extremes of the observed
\phigc\ range. This would seem like a plausible explanation for the
measured gradient in \phigc, as the data-points in
Fig.\ \ref{fig:azgrad} are few and poorly sampled at the extremes of
\phigc.

\begin{figure*}
  \centering
  \includegraphics[width=8.cm]{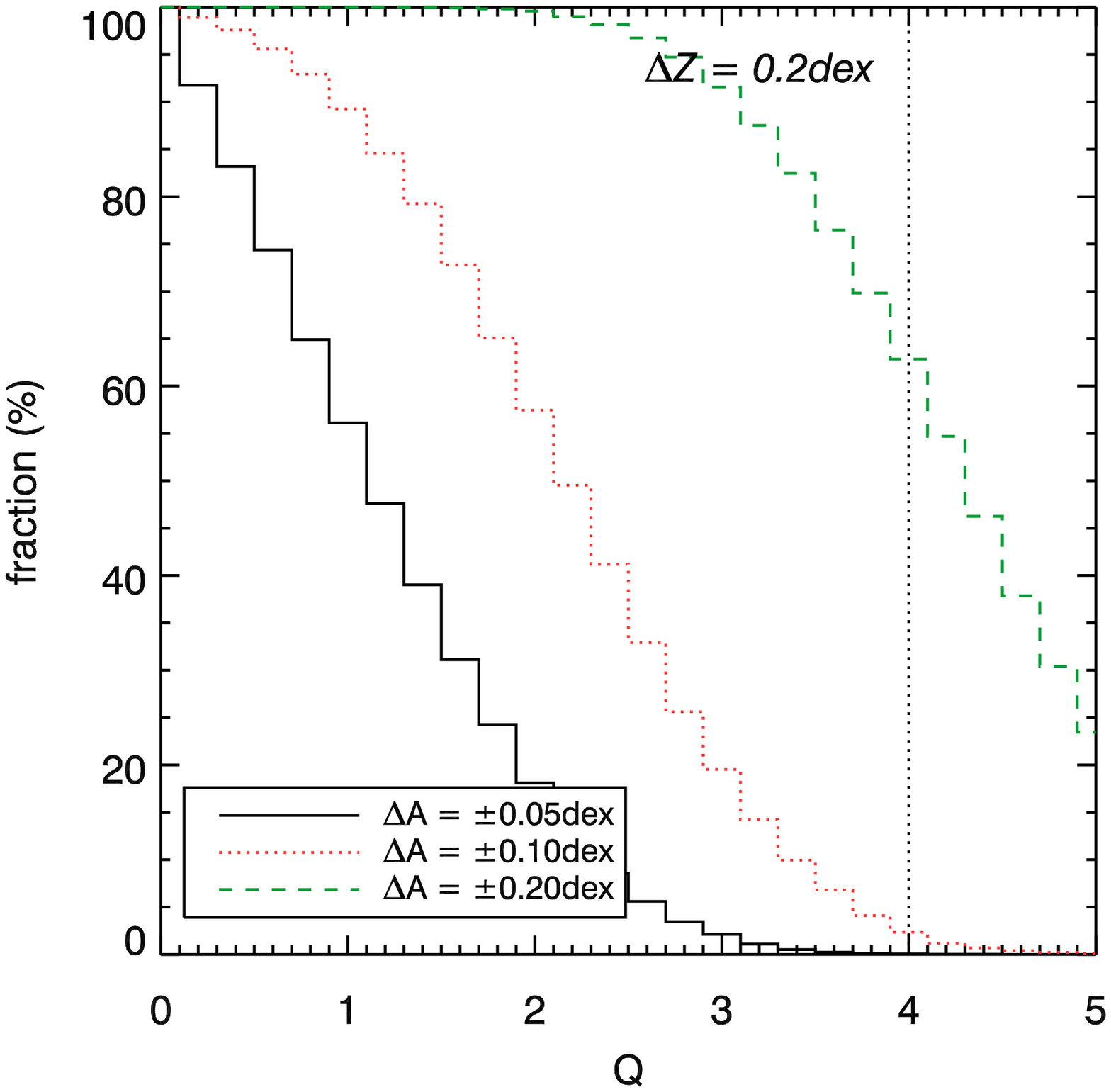}
  \includegraphics[width=8.cm]{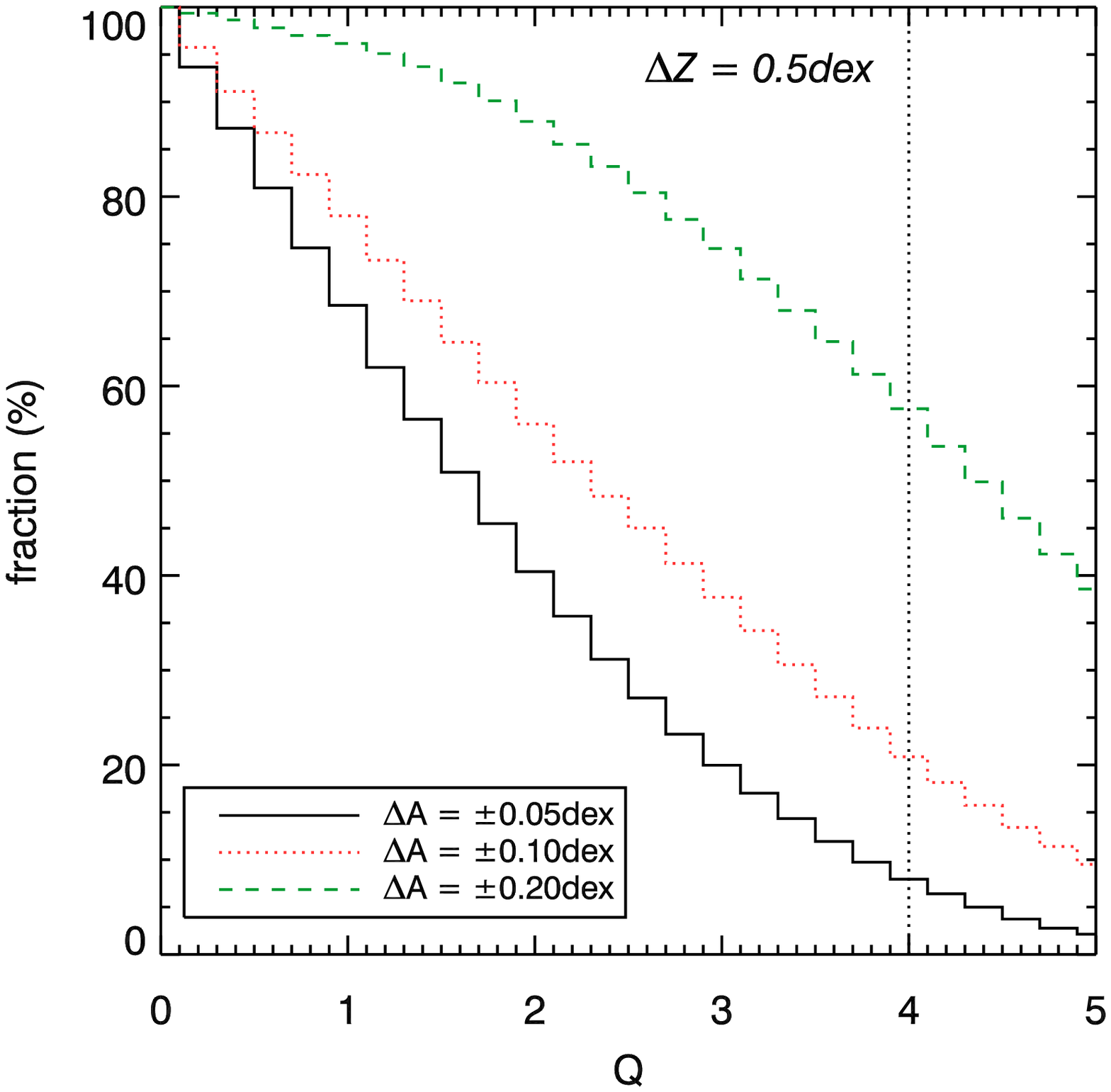}
  \caption{The probability of obtaining a gradient with significance
  $Q$ ($\equiv g / \sigma g$ ) for three different values of $\Delta
  A$, which is defined as systematic difference in the measured
  abundance of an element when determined from two different
  methods. We plot simulations for two different {\it intrinsic}
  abundance scatters: $\Delta Z = \pm$0.2dex (left) and $\Delta Z =
  \pm$0.5dex (right). The significance of our measured azimuthal
  gradients in O and Si, $Q=4$, is indicated by the dotted line. Each
  plot shows that, for larger systematic differences, the probability
  of artificially inferring a gradient is higher. Meanwhile,
  comparison between the two figures shows that gradients are more
  likely to be inferred if the intrinsic scatter is larger.\\ }
  \label{fig:sys}
\end{figure*}

To investigate quantitatively the likelihood of this scenario, we have
performed monte-carlo simulations to generate random abundance levels
at the {\it same} sampling in $\cos$($\phi'$)-space as the data
plotted in Fig.\ \ref{fig:azgrad}. For each simulation, we generated
data with uniformly random distribution with a peak-to-peak amplitude
of 0.85dex -- similar to that of the observed data. We set the
uncertainty in each value of $A(X)$ to the same as in the observations
at that value of $\cos$($\phi'$)\footnote{Setting all measurement
  uncertainties to $\pm$0.15dex had no effect on the results of our
  analysis.}. We calculated the gradient $g$ and uncertainty $\sigma
g$ of the simulated data, and measured the significance of the
gradient $Q \equiv g/\sigma g$. The simulation was repeated 10$^{5}$
times and the frequency of obtaining a measured gradient with
significance $Q \ge 4$ was recorded.

We found that the likelihood of measuring a gradient at the $Q=4$
significance level in the O and Si data was low, $\sim$14\%. As it may
be argued that the two RSGC measurements may not be independent of one
another, we then repeated the experiment, this time forcing the two
RSGC data-points to the same value. This time, the probabilities of
obtaining a $Q=4$ gradient rose to $\approx$23\% for each of O and
Si. We repeated this analysis, this time generating random $A(X)$
values from a guassian distribution with standard deviation $\sigma A
= 0.3$, the observed spread of the Si abundances. This time, we found
that the probability of obtaining $Q=4$ was 22\%, or 30\% when the
RSGC points were forced to the same value.

From this analysis, we conclude that the measured gradients in Si and
O are unlikely to be a result of random sampling. That is to say, if
large azimuthal abundance dispersions exist in the inner disk, the
likelihood of artificially inferring the gradients we measure from
random sampling is $\la$30\%.

\paragraph{Systematic offsets between analysis methods --} Another
explanation for the observed gradient may be that the abundances in
the inner disk are uniform, but a systematic offset exists between
abundances measured using different methods. For example, Si
abundances measured from Cepheids may yield values which are
systematically higher than those measured from RSGs, even though the
{\it true} abundances of the two objects are the same. Indeed, one may
argue from the bottom panel of Fig.\ \ref{fig:azgrad} that the Si
abundances of the Cepheids are constant to within the errors, as are
the two RSG clusters, with the two classes of objects offset from one
another by $\sim$0.5dex.

We performed a quantitative investigation of this, again using
monte-carlo simulations. We assumed that the abundances $Z$ in the
inner disk are uniform, with an intrinsic peak-to-peak scatter of
$\Delta Z$. We randomly generated abundance values at the same
distribution in $\cos(\phi')$ as the data in Fig.\ \ref{fig:azgrad}. We
then artificially raised the abundances of the cepheid points by
$\Delta A$, and decreased the RSGC points by the same quantity. That
is, a value of $\Delta A = \pm 0.2$ separates the two classes of
object by 0.4dex in abundance space. We again forced the RSGC
abundaces to be the same, as these measurements may not be independent
of one another. The abundances of the B~stars were randomized, with a
distribution of $\pm$0.4dex about the mean value, similar to the data
in Fig.\ \ref{fig:azgrad}. At the end of each simulation, the gradient
$g$ and its significance $Q$ was measured, and the experiment repeated
10$^{5}$ times.

In Fig.\ \ref{fig:sys}, we plot the likelihood of measuring a
statistically significant gradient for three different values of
$\Delta A$, using the same $\cos(\phi')$ values as those in the bottom
panel of Fig.\ \ref{fig:azgrad}. We then reproduce this analysis for
two different {\it intrinsic} abundance scatters, $\Delta Z$. For
example, the right-hand panel of Fig.\ \ref{fig:sys} shows the
fraction of simulations for which a gradient with significance $Q$ is
measured for data with a peak-to-peak abundance scatter of $\Delta Z
=$0.5dex. In this panel, for a systematic offset $\Delta A =
\pm0.1$dex -- the same as quoted uncertainties in the data -- the
probability of inferring a $Q>4$ gradient is 20\%. For an offset of
$\Delta A = \pm0.2$, which is the value typically quoted as being the
absolute uncertainty on any abundance measurement, the probability is
raised to 50-60\%. 

From this analysis, we conclude that it is possible that the azimuthal
abundance gradient we measure is due to systematic offsets between
differing measurement techniques. However, for a 50\% probability of
obtaining the $Q=4$ gradients we observe, the offset between the RSGCs
and the Cepheids must be large, $\sim$0.4dex (i.e.\ $\Delta A = \pm
0.2$). We note that in the Galactic Centre, we found good agreement
(better than $\pm$0.2dex) between our analysis of RSGs and that of
other classes of objects such as Luminous Blue Variables
\citep{Najarro08,RSGGC}.

\paragraph{Real trends --} Having studied two scenarios that could
lead us to artificially measure a gradient where none existed, we now
discuss the possibility that the measured azimuthal abundance
variations are real. If such behaviour exists in our Galaxy, then it
is reasonable to assume that it should also be present in external
galaxies. In the following section, we investigate the presence of
azimuthal abundance variations in the face-on spiral NGC~4736.

\begin{figure}[t]
  \centering
  \includegraphics[width=8.cm]{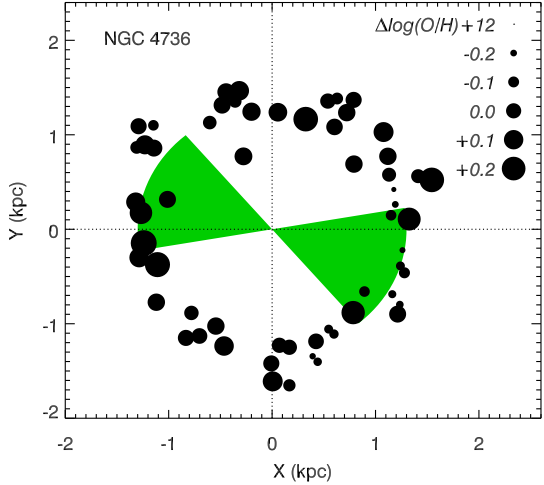}
  \caption{Azimuthal [O/H] variations in NGC~4736. Symbol sizes
    represent the magnitude of the abundance levels in the various
    \hii regions that surround the central region, as indicated by the
    legend. The shaded region denotes the rough extent and position
    angle (110$\pm$30\degr) of the galaxy's bar. }
  \label{fig:ngc4736}
\end{figure}

\subsubsection{Case-study: NGC~4736 ($\equiv$ M94) }
In looking for an azimuthal abundance gradient in an external galaxy,
we have chosen this object for several reasons. Firstly, while not
having the exact same type as our own Galaxy, it is a spiral with a
central bar, and is relatively face-on \citep{Block94}. Secondly, it
has a ring of \hii\ regions in the inner disk, which are well sampled
in azimuth \citep{Lynds74}. Thirdly, oxygen abundances of these \hii\
regions are present in the literature: \citet{M-B97} studied the \hii\
region ring with narrow-band filters to measure the fluxes of certain
diagnostic lines in the optical, and used them to derive
abundances. However, the authors did not use the abundances to study
azimuthal variations, only the presence of a radial gradient. This
study utilized a homogeneous methodology, and so we do not have to
take into account systematic errors induced by differing analysis
techniques.

\begin{figure}[t]
  \centering
  \includegraphics[width=8.cm]{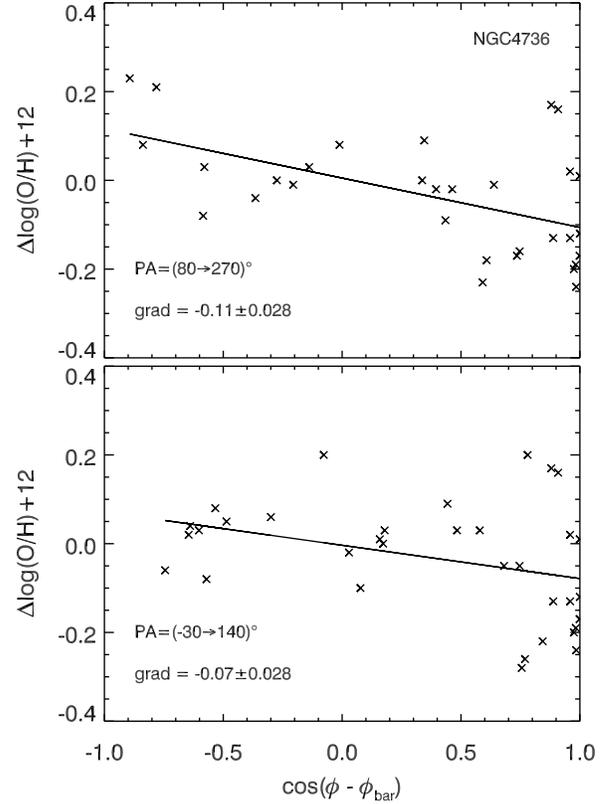}
  \caption{The abundance ratio [O/H] as a function of azimuthal angle,
    with the coordinate system rotated such that the central bar
    ($\phi_{/rm bar}$=110$\pm$30\degr) is oriented at $\cos(\rm \phi')
    = 1$. In the top panel, we plot [O/H] versus azimuthal angle in
    the clockwise direction away from the bar, plotting those points
    with $\phi >$80\degr (=$ \phi_{\rm bar}-\sigma \phi_{\rm
      bar}$). Similarly, in the bottom panel we [O/H] as a function of
    $\cos(\phi)$ in the anti-clockwise direction, plotting those
    points with $\phi < 140$\degr ($= (\phi_{\rm bar}+\sigma \phi_{\rm
      bar}$) }
  \label{fig:ngc4736_sinphi}
\end{figure}

In Fig.\ \ref{fig:ngc4736} ({\it left}) we plot the locations of the
\hii\ regions in the $X-Y$ plane, corrected for the galaxy's
inclination. The size of the plotting symbols indicates the magnitude
of the abundance of each \hii\ region with respect to the global
average. To the eye, it seems that the abundances in the lower-right
quadrant of the figure tend to be lower than average; while at the
left edge of the ring at $(X,Y) \approx (-1,0)$ the abundances are
highest. Interestingly, the minimum in abundance levels corresponds
roughly with the end of the bar \citep[$\phi_{\rm bar}$ =
  110$\pm$30\degr, indicated by the shaded region in
  Fig.\ \ref{fig:ngc4736} -- ][]{Block94,W-B00}. This is similar to
the trend seen in our own Galaxy, where the abundance minimum is found
in the RSG clusters located at the end of the Galaxy's Bar.

In Fig.\ \ref{fig:ngc4736_sinphi}, we plot the abundance levels as a
function of the cosine of the azimuthal angle. Again, as in
Fig.\ \ref{fig:azgrad} we have rotated the coordinate system to such
that the minimum abundances correspond to $\cos(\phi) = 1.0$. For this
rotation angle we adopt the position angle of the bar,
110$\pm$30\degr. The top panel of Fig.\ \ref{fig:ngc4736_sinphi} shows
the abundance trend as a function of $\cos(\phi)$ in the clockwise
direction from the lower limit of the bar's orientation (110\degr -
30\degr = 80\degr). Similarly, in the bottom panel we show the
abundance trend in the anti-clockwise direction up to the upper limit
of the bar's orientation (140\degr). We opt to plot the abundance
variations in the two directions separately, as the trends in each
direction are not necessarily similar. 

In each panel of Fig.\ \ref{fig:ngc4736_sinphi} we make a linear fit
to the [O/H] distribution and indicate on the plot the measured
gradient and its uncertainty. The maximum position angle that we have
chosen to plot in each panel, $\phi_{bar} \pm$160\degr, was somewhat
arbitrarily chosen to illustrate a wide range of azimuth as well as
include a sufficiently large number of data-points, though we note
that the range chosen does not significantly affect $Q$-values of the
measured gradients.

From the two panels of Fig.\ \ref{fig:ngc4736_sinphi} we see that
abundance gradients are measured at the $Q=4$ (clockwise) and $Q=2$
(anti-clockwise) levels. Though tentative, this does provide some
support for the idea of localised azimuthal abundance variations in
the inner disk of NGC~4736. We again state that, as in our analysis of
the inner Galaxy, these `gradients' are more likely to be indicative
of systematic undulations in abundance levels rather than true
sinusoidal variations with azimuthal angle.

\subsubsection{Bars, and azimuthal abundance variations}
To summarize the results of the previous section, we find it unlikely
that the azimuthal abundance patterns we observe in the inner Galaxy
are an artifact of random sampling. We cannot rule out that the
observed variations result from systematic offsets between different
methodologies. However, in using data from a homogeneous study of the
external galaxy NGC~4736 where we are able to make a differential
analysis, we find tentative evidence for similar abundance
variations. In each case, the lowest abundances were observed at the
end of the central bar.

Given these apparent azimuthal variations in abundances seen in both
our Galaxy and NGC~4736, we now discuss how such could patterns
arise. It is known that there is a link between abundance trends and
the presence of bars -- radial abundance gradients tend to be flatter
in barred spirals than in unbarred spirals
\citep{Zaritsky94,Mar-Roy94}. Indeed, the observed abundance gradient
in our Galaxy is typical of barred spirals, and much shallower than
unbarred spirals. The common explanation for this is that radial gas
motions induced by the bar potential smooth-out any pre-existing
abundance gradient on timescales of a few $\times$100Myr. However,
this can introduce significant azimuthal variations, especially
between the spiral arms and the inter-arm regions, due to intense but
patchy star-formation \citep{Friedli94,F-B95}.

In NGC~4736, kinematic studies have shown that the \hii\ ring is
coincident with the outer Lindblad resonance of the bar and the inner
Lindblad resonance of the central bulge. The ring itself is likely a
result of radial inward gas motions outside of this resonance zone and
outward motions inside \citep{W-B00,MT04}. The ensuing star-formation
within the ring, as the colliding material is shocked, has been shown
to be patchy from CO and \halpha\ observations \citep{W-B00}. As the
timescale for chemical enrichment by massive stars is comparable to
the dynamical timescales in the region of the bar (few
$\times$100Myr), it is conceivable that notable fluctuations in the
azimuthal abundances of the $\alpha$-elements would
survive. Similarly, in our Galaxy, azimuthal abundance variations
could be produced by intense but patchy star-formation in the region
of the Bar's outer Lindblad resonance at \rgc $\approx$4\,kpc. Indeed,
this starburst episode could be responsible for creating the RSG
clusters. However, it is unclear how this effect would show itself in
the Fe abundances, as the enrichment timescale for this element is
longer ($\sim 10^{8} - 10^{9}$yr). Clearly, the abundances in the
inner Galactic disk require further study, and would benefit from a
homogenous study such as that of NGC~4736 by \citet{M-B97}. Massive
young clusters such as RSGC1 and RSGC2, from which abundances of
several elements may be determined in the near-IR, may represent the
best tool with which to do this. Though known examples of such
clusters are currently low in number, searches of infrared Galactic
plane surveys have revealed many more candidates, and the number of
known young massive clusters in the inner Galaxy looks set to rise in
the next few years.

\section{Conclusions} \label{sec:conc}
We have used high-resolution near-infrared spectroscopy and detailed
spectral synthesis to determine the chemical abundances of the two
Scutum Red Supergiant clusters, RSGC1 and RSGC2, which are located in
the inner Galaxy close to the end of the Bar. Our results can be
summarized as follows:

\begin{itemize}
  \item The stars atmopsheres are carbon-depleted, and the average
  levels of depletion agree well with the predictions of rotating
  single-star models. Non-rotating models predict C/Fe surface
  abundances which are higher than observed. 
  \item The average iron abundances (Fe/H) are found to be sub-solar
  by 0.2-0.3dex. This is in apparent conflict with the Fe abundances
  of objects at similar Galactocentric distances, which have typically
  been found to be $\sim$0.5dex higher.
  \item The abundances of the $\alpha$-elements are also
  sub-solar. The ratio of $\alpha$/Fe is fully consistent with solar,
  as is typical for objects in the thin disk. 
\end{itemize}

To explain the apparently unusual abundances of the RSG clusters with
respect to other objects in the inner disk, we have collected several
abundance measurements of different classes of object in the
literature. It appears that large azimuthal abundance variations exist
in the inner disk, as is predicted by theory for the secular evolution
of galaxies with bars. Further, we find that the objects in the inner
disk with strongly super-solar abundances are typically located at
negative Galactic longitude $l$, while the RSG clusters are located at
positive $l$. We find tentative evidence for large-scale azimuthal
abundance variations in the data, significant at the
4-5$\sigma$-level. The presence of these variations can only be
explained away if abundance studies of different types of object have
significant systematic offsets from one another. In a differential
analysis study of the external barred galaxy NGC~4736 using
homogeneous methodology, we again find tentative evidence for
large-scale azimuthal abundance fluctuations. We suggest that these
results for both our Galaxy and NGC~4736 are consistent with the
predictions of the Friedi \& Benz model for chemical evolution in
barred spirals, whereby the bar potential which causes a pile-up of
material at the outer Lindblad resonance results in intense but patchy
star-formation in the inner region of the disk. As the dynamical
timescale is comparable to the timescale for chemical enrichment,
large azimuthal abundance variations can form.

\acknowledgments Acknowledgements: For many useful discussions, we
would like to thank Nate Bastian, Bob Benjamin, Paul Crowther, Katia
Cunha, John Eldridge, Maria Messineo, and Rob Kennicutt. We thank
Bengt Gustafsson for providing us with his group's molecular line-list
for CN. We also thank the anonymous referee for several comments and
suggestions which helped us improve the paper. The material in this
work is supported by NASA under award NNG~05-GC37G, through the
Long-Term Space Astrophysics program. This research was performed in
the Rochester Imaging Detector Laboratory with support from a NYSTAR
Faculty Development Program grant. The data presented here were
obtained at the W.\ M.\ Keck Observatory, which is operated as a
scientific partnership among the California Institute of Technology,
the University of California, and the National Aeronautics and Space
Administration. The Observatory was made possible by the generous
financial support of the W.\ M.\ Keck Foundation. This research has
made use of IDL software packages, and the GSFC IDL library.

\bibliographystyle{/fat/Data/bibtex/apj}
\bibliography{/fat/Data/bibtex/biblio}

\appendix {\bf Appendix:} The chemical abundances of the stars observed in each
cluster. Star identification numbers are taken from \citet{Figer06}
and \citet{RSGC2paper}

\begin{deluxetable}{lcccccccc}
\tabletypesize{\scriptsize}
\tablecaption{Elemental abundances for each star studied. Abundances
  for each element $X$ are expressed in the form $A(X) = \log (X/{\rm
    H}) +12$. For reference, the solar abundances derived by Asplund et
  al.\ (2005) are also shown. \label{tab:data}}
\tablewidth{0pt}
\tablehead{
\colhead{Star} &
\colhead{$\rm T_{eff}$}&
\colhead{$\rm A(Fe)$}&
\colhead{$\rm A(O)$}&
\colhead{$\rm A(Si)$}&
\colhead{$\rm A(Mg)$}&
\colhead{$\rm A(Ca)$}&
\colhead{$\rm A(Ti)$}&
\colhead{$\rm A(C)$}
}
\startdata
\multicolumn{9}{l}{\smallskip Solar Abundances (Asplund et al. 2005) }\\
 -- & --  &       7.45  &       8.66  &       7.51  &       7.53  &       6.31  &       4.90  &       8.39 \\
 -- & --  & $\pm$ 0.05  & $\pm$ 0.05  & $\pm$ 0.04  & $\pm$ 0.09  & $\pm$ 0.04  & $\pm$ 0.06  & $\pm$ 0.05 \\
\hline \\
\multicolumn{9}{l}{\smallskip RSGC1 Abundances}\\
 1 & 3600 &       7.36 &       8.58 &       7.21 &       7.57 &       6.16 &       4.98 &       7.82 \\
 -- & --  & $\pm$ 0.09 & $\pm$ 0.08 & $\pm$ 0.14 & $\pm$ 0.12 & $\pm$ 0.07 & $\pm$ 0.10 & $\pm$ 0.09 \\
 2 & 3600 &       7.30 &       8.59 &       7.25 &       7.28 &       6.19 &       4.72 &       7.92 \\
 -- & --  & $\pm$ 0.08 & $\pm$ 0.09 & $\pm$ 0.14 & $\pm$ 0.12 & $\pm$ 0.07 & $\pm$ 0.10 & $\pm$ 0.09 \\
 3 & 3400 &       7.40 &       8.58 &       7.30 &       7.48 &       6.24 &       5.02 &       7.92 \\
 -- & --  & $\pm$ 0.09 & $\pm$ 0.13 & $\pm$ 0.15 & $\pm$ 0.13 & $\pm$ 0.08 & $\pm$ 0.12 & $\pm$ 0.09 \\
 4 & 3800 &       7.35 &       8.76 &       7.32 &       7.36 &       6.28 &       5.02 &       7.92 \\
 -- & --  & $\pm$ 0.09 & $\pm$ 0.07 & $\pm$ 0.11 & $\pm$ 0.12 & $\pm$ 0.07 & $\pm$ 0.08 & $\pm$ 0.09 \\
 5 & 3600 &       7.29 &       8.63 &       7.22 &       7.28 &       6.20 &       4.92 &       7.97 \\
 -- & --  & $\pm$ 0.08 & $\pm$ 0.10 & $\pm$ 0.14 & $\pm$ 0.12 & $\pm$ 0.07 & $\pm$ 0.10 & $\pm$ 0.09 \\
 6 & 3400 &       7.41 &       8.58 &       7.35 &       7.64 &       6.29 &       5.02 &       8.02 \\
 -- & --  & $\pm$ 0.08 & $\pm$ 0.12 & $\pm$ 0.15 & $\pm$ 0.13 & $\pm$ 0.08 & $\pm$ 0.12 & $\pm$ 0.09 \\
 7 & 3600 &       7.27 &       8.54 &       7.25 &       7.28 &       6.12 &       4.82 &       7.92 \\
 -- & --  & $\pm$ 0.08 & $\pm$ 0.09 & $\pm$ 0.14 & $\pm$ 0.12 & $\pm$ 0.07 & $\pm$ 0.09 & $\pm$ 0.09 \\
 8 & 3600 &       7.32 &       8.61 &       7.25 &       7.27 &       6.18 &       4.92 &       8.02 \\
 -- & --  & $\pm$ 0.08 & $\pm$ 0.09 & $\pm$ 0.14 & $\pm$ 0.12 & $\pm$ 0.07 & $\pm$ 0.10 & $\pm$ 0.09 \\
 9 & 3400 &       7.37 &       8.53 &       7.25 &       7.54 &       6.19 &       4.92 &       7.92 \\
 -- & --  & $\pm$ 0.09 & $\pm$ 0.11 & $\pm$ 0.15 & $\pm$ 0.13 & $\pm$ 0.08 & $\pm$ 0.10 & $\pm$ 0.09 \\
10 & 3600 &       7.31 &       8.72 &       7.35 &       7.48 &       6.22 &       5.02 &       8.02 \\
 -- & --  & $\pm$ 0.09 & $\pm$ 0.10 & $\pm$ 0.14 & $\pm$ 0.12 & $\pm$ 0.08 & $\pm$ 0.11 & $\pm$ 0.09 \\
11 & 3600 &       7.29 &       8.57 &       7.35 &       7.38 &       6.21 &       4.92 &       8.02 \\
 -- & --  & $\pm$ 0.08 & $\pm$ 0.10 & $\pm$ 0.14 & $\pm$ 0.12 & $\pm$ 0.07 & $\pm$ 0.10 & $\pm$ 0.09 \\
12 & 3800 &       7.38 &       8.70 &       7.53 &       7.53 &       6.26 &       4.92 &       8.02 \\
 -- & --  & $\pm$ 0.09 & $\pm$ 0.07 & $\pm$ 0.11 & $\pm$ 0.12 & $\pm$ 0.06 & $\pm$ 0.07 & $\pm$ 0.09 \\
13 & 3800 &       7.27 &       8.68 &       7.25 &       7.28 &       6.18 &       4.72 &       7.82 \\
 -- & --  & $\pm$ 0.07 & $\pm$ 0.07 & $\pm$ 0.11 & $\pm$ 0.12 & $\pm$ 0.07 & $\pm$ 0.08 & $\pm$ 0.09 \\
14 & 3600 &       7.25 &       8.54 &       7.35 &       7.38 &       6.13 &       4.92 &       7.92 \\
 -- & --  & $\pm$ 0.08 & $\pm$ 0.10 & $\pm$ 0.14 & $\pm$ 0.12 & $\pm$ 0.07 & $\pm$ 0.10 & $\pm$ 0.09 \\
\hline \\
\multicolumn{9}{l}{\smallskip RSGC2 Abundances}\\
 2 & 3600 &       7.04 &       8.37 &       7.05 &       7.08 &       6.01 &       4.73 &       7.72 \\
 -- & --  & $\pm$ 0.08 & $\pm$ 0.09 & $\pm$ 0.13 & $\pm$ 0.12 & $\pm$ 0.06 & $\pm$ 0.08 & $\pm$ 0.09 \\
 3 & 3600 &       7.22 &       8.60 &       7.24 &       7.28 &       6.11 &       4.92 &       7.92 \\
 -- & --  & $\pm$ 0.08 & $\pm$ 0.10 & $\pm$ 0.14 & $\pm$ 0.12 & $\pm$ 0.07 & $\pm$ 0.10 & $\pm$ 0.09 \\
 5 & 3600 &       7.29 &       8.54 &       7.25 &       7.25 &       6.18 &       4.92 &       7.92 \\
 -- & --  & $\pm$ 0.08 & $\pm$ 0.10 & $\pm$ 0.14 & $\pm$ 0.12 & $\pm$ 0.07 & $\pm$ 0.10 & $\pm$ 0.09 \\
 6 & 3600 &       7.30 &       8.60 &       7.50 &       7.53 &       6.21 &       4.92 &       7.82 \\
 -- & --  & $\pm$ 0.08 & $\pm$ 0.10 & $\pm$ 0.14 & $\pm$ 0.12 & $\pm$ 0.07 & $\pm$ 0.10 & $\pm$ 0.09 \\
 8 & 3600 &       7.30 &       8.70 &       7.45 &       7.38 &       6.23 &       4.92 &       8.02 \\
 -- & --  & $\pm$ 0.08 & $\pm$ 0.08 & $\pm$ 0.13 & $\pm$ 0.12 & $\pm$ 0.08 & $\pm$ 0.11 & $\pm$ 0.09 \\
 9 & 3600 &       7.30 &       8.58 &       7.35 &       7.38 &       6.21 &       4.92 &       7.92 \\
 -- & --  & $\pm$ 0.08 & $\pm$ 0.10 & $\pm$ 0.14 & $\pm$ 0.12 & $\pm$ 0.07 & $\pm$ 0.10 & $\pm$ 0.09 \\
10 & 3600 &       7.27 &       8.54 &       7.35 &       7.22 &       6.15 &       4.92 &       7.92 \\
 -- & --  & $\pm$ 0.08 & $\pm$ 0.10 & $\pm$ 0.14 & $\pm$ 0.12 & $\pm$ 0.07 & $\pm$ 0.10 & $\pm$ 0.09 \\
11 & 3600 &       7.26 &       8.64 &       7.35 &       7.31 &       6.21 &       4.92 &       7.92 \\
 -- & --  & $\pm$ 0.08 & $\pm$ 0.10 & $\pm$ 0.14 & $\pm$ 0.12 & $\pm$ 0.07 & $\pm$ 0.10 & $\pm$ 0.09 \\
14 & 3600 &       7.29 &       8.62 &       7.35 &       7.38 &       6.21 &       4.92 &       7.82 \\
 -- & --  & $\pm$ 0.08 & $\pm$ 0.09 & $\pm$ 0.14 & $\pm$ 0.12 & $\pm$ 0.07 & $\pm$ 0.10 & $\pm$ 0.09 \\
15 & 3600 &       7.26 &       8.56 &       7.35 &       7.38 &       6.16 &       4.92 &       7.82 \\
 -- & --  & $\pm$ 0.08 & $\pm$ 0.10 & $\pm$ 0.14 & $\pm$ 0.12 & $\pm$ 0.07 & $\pm$ 0.10 & $\pm$ 0.09 \\
21 & 3600 &       7.32 &       8.54 &       7.35 &       7.38 &       6.13 &       4.92 &       7.82 \\
 -- & --  & $\pm$ 0.08 & $\pm$ 0.10 & $\pm$ 0.14 & $\pm$ 0.12 & $\pm$ 0.07 & $\pm$ 0.10 & $\pm$ 0.09 \\
29 & 3800 &       7.29 &       8.59 &       7.45 &       7.49 &       6.20 &       4.92 &       7.92 \\
 -- & --  & $\pm$ 0.08 & $\pm$ 0.08 & $\pm$ 0.11 & $\pm$ 0.12 & $\pm$ 0.07 & $\pm$ 0.08 & $\pm$ 0.09 \\
\enddata
\end{deluxetable}

\end{document}